\documentclass[conference]{IEEEtran}
\IEEEoverridecommandlockouts

\usepackage{times}
\usepackage{graphicx}
\usepackage{cite,xspace,url,setspace}
\usepackage{float, rotating}
\usepackage[cmex10]{amsmath}
\usepackage{amsthm}
\usepackage{amsmath,amsfonts}
\usepackage{algorithmic,booktabs}
\usepackage{array}
\usepackage{bm}
\usepackage{hhline}

\usepackage[algoruled, linesnumbered]{algorithm2e}

\usepackage[tight,footnotesize]{subfigure}
\usepackage{url,caption, color, enumerate}

\title{Deep Reinforcement Learning for Power Control in Next-Generation WiFi Network Systems\thanks{DISTRIBUTION STATEMENT A.  Approved for public release.}
}

\author{\IEEEauthorblockN{
		Ziad El Jamous$^{1}$,
		Kemal Davaslioglu$^2$, and
		Yalin E. Sagduyu$^{3}$
	}
	\IEEEauthorblockA{$^{1}$BlueHalo, Rockville, MD, USA}
	\IEEEauthorblockA{$^{2}$University Technical Services, Inc., Greenbelt, MD, USA}
	\IEEEauthorblockA{$^{3}$National Security Institute, Virginia Tech, Arlington, VA, USA}
	\IEEEauthorblockA{Email: ziad.eljamous@bluehalo.com, kemal@ut-services.com, ysagduyu@vt.edu}
}
\begin{document}
	\newcommand{\argmax}{\arg\!\max}
	\maketitle
	
	\begin{abstract}
		This paper presents a deep reinforcement learning (DRL) solution for power control in wireless communications, describes its embedded implementation with WiFi transceivers for a WiFi network system, and evaluates the performance with high-fidelity emulation tests. In a multi-hop wireless network, each mobile node measures its link quality and signal strength, and controls its transmit power. As a model-free solution, reinforcement learning allows nodes to adapt their actions by observing the states and maximize their cumulative rewards over time. For each node, the state consists of transmit power, link quality and signal strength; the action adjusts the transmit power; and the reward combines energy efficiency (throughput normalized by energy consumption) and penalty of changing the transmit power. As the state space is large, Q-learning is hard to implement on embedded platforms with limited memory and processing power. By approximating the Q-values with a deep Q-network (DQN), DRL is implemented for the embedded platform of each node combining an ARM processor and a WiFi transceiver for 802.11n. Controllable and repeatable emulation tests are performed by inducing realistic channel effects on RF signals. Performance comparison with benchmark schemes of fixed and myopic power allocations shows that power control with DRL provides major improvements to energy efficiency and throughput in WiFi network systems.

	\end{abstract}
	\section{Introduction}
	Machine learning (ML) provides powerful means to learn from rich data representations and solve complex tasks. Recent algorithmic and computational advances have supported the broad use of deep neural networks for diverse applications ranging from computer vision to natural language processing. As high-fidelity training data with labels is not always readily available, reinforcement learning (RL) has emerged as a viable solution for taking actions to maximize a cumulative reward over a time horizon while interacting with and learning from the environment. As a model-free RL algorithm, Q-learning aims to learn the value of an action in a particular state, namely the Q-value. However, the number of Q-value evaluations grows with the state and action space, making the maintenance of Q-tables computationally infeasible for a large-scale problem as well as an embedded implementation (as needed for edge computing in wireless communication applications). Deep reinforcement learning (DRL)  has emerged as a computationally efficient solution by approximating the Q-values by training a deep Q-network (DQN).
	
	As the spectrum data associated with wireless communication networks is highly complex due to network topology, channel, interference, and traffic effects, deep learning (DL) finds diverse sets of wireless applications including RF signal classification, waveform design, and security \cite{WirelessDL}. Wireless networks involve dynamic effects with temporal correlations such as those induced by network mobility and waveform patterns. Therefore, RL has the strong potential in supporting wireless communication systems to learn from the network environment and optimize the network performance over time without relying on the availability of training data or any other prior knowledge on the underlying model to learn \cite{WirelessDRL}. The RL solutions have been studied for general communication systems with various tasks including channel access \cite{DRLMAC1, DRLMAC2, DRLMAC3, DRLMAC4, DRLMAC5, DRLMAC6, DRLMAC7, DRLMAC8, DRLMAC9, DRLMAC10, DRLMAC11} and power control \cite{DRLPower1, DRLPower2, DRLPower3, DRLPower4, DRLPower5, DRLPower6}. The typical approach in previous works has been to use analytical models and evaluate the performance via theoretical evaluations or simulations. Instead, the real protocol stack of WiFi communications is used in this paper to derive the states, actions, and rewards of DRL (that is needed to tackle the complexity of the underlying optimization problem), and high-fidelity experiments with real radios are performed to show the benefits of DRL for WiFi network systems. 
	
	In this paper, a multi-hop wireless network system is considered, where each node is equipped with an embedded platform and a WiFi (in particular, 802.11n) transceiver. Each node may act as the source, the destination, or the relay for network packet traffic. With the emergence of smart WiFi systems such as in WiFi 7 \cite{WiFi7}, it is necessary to optimize the communication functionalities for WiFi to meet the growing demand of emerging applications, including extended reality (XR), gaming, video conferencing and streaming. These applications require high-rate communications and benefit from energy efficiency with prolonged network lifetime (as needed for battery-operated networked devices). 
	
	There are many configuration parameters for next-generation WiFi systems that involve complex dependencies between parameters and their joint optimization especially in dense deployments and coexistence in shared bands. ML has been considered as viable means to optimize next-generation WiFi systems \cite{WiFiML}. Potential scenarios include 5G/6G-WiFi convergence, Multiple Radio Access Technology (multi-RAT) including WiFi, and Internet of Things (IoT) devices connected with WiFi (potentially in multi-hop network setting such as in smart warehouses, where multiple hops are needed to reach IoT devices over an extended area) and mesh WiFi networks such as Google Nest and Amazon Echo. All these applications can benefit from improvements in both energy consumption (that translates to longer battery and network lifetime) and throughput \cite{GreenIoT}. 
	DL has been used for adaptation of channel access mechanisms in WiFi to spectrum dynamics \cite{DeepWiFi} and DRL has been used for coexistence of WiFi with cellular communications \cite{WiFiSharing}. 
	
	The backoff mechanism of carrier-sense multiple access with collision avoidance (CSMA/CA) as used in 802.11 protocols generates temporal correlations in the channel access pattern. Also, network mobility induces temporal correlations in observed spectrum data. Note that no training data is assumed to be available in advance for the use of supervised ML for the adaptation of WiFi communications to spectrum dynamics. Similarly, no prior model is assumed to be available to formulate an optimization problem in advance to solve. Therefore, RL is considered in this paper for  WiFi nodes to adapt themselves to protocol and network dynamics, and optimize their energy efficiency which is a critical concern especially for battery-operated devices. 
	In addition, a distributed RL approach is pursued without considering a centralized node to prevent communication and computation bottlenecks and avoid a single point of failure.
	
	In this RL formulation, the states are the transmit power, link quality and signal strength that are either maintained or measured by each node. The actions adjust the transmit power. As the corresponding Q-table is large, a DRL solution is used in this paper and implemented on the embedded platform of each node. The objectives of maximizing the network throughput and minimizing the energy consumption (that is critical for edge devices operating on battery power) are integrated in an energy efficiency objective. A reward function is built to combine the energy efficiency with the penalty on changing the transmit power excessively (to reduce the extra processing burden and avoid performance fluctuations). 
	
	For the system-level assessment, the proposed DRL solution for power control is implemented on the ARM processor of each node's embedded platform that connects to an Alfa AWUS036NHA USB wireless adapter for 802.11n communications at the 2.4~GHz band (while the algorithm is independent of frequency band). The performance is measured in terms of energy efficiency and throughput. 
	
	Performance evaluation is carried out with high-fidelity emulation tests by employing real transceivers and embedded implementation of real network protocol stack (including the proposed power control solution as well as other network protocols like routing for multi-hop network operations) and real packet traffic. Channel effects among transceivers are emulated with a network channel emulator. This evaluation approach is more realistic than simulations that typically use simplistic models of physical layer that ignore effects such
	as nonlinearity, filtering, and inter modulation that are
	caused by hardware. An alternative is making over the air (OTA) transmissions in a traditional testbed environment. However, channel effects change over time and OTA tests cannot guarantee repeatable and controllable evaluation of the proposed scheme and benchmark schemes under the exact same scenario corresponding to the same network topology/mobility and channel effects \cite{emulationtestbed, Magazine}. 
	
	For the emulation tests in this paper, RFnest \cite{RFnest} is used as the network channel emulator. WiFi transceivers are connected to RFnest with cables. RFnest controls channel effects on RF signals according to  the network topology and mobility pattern, and allows channel sensing and interference, if any, among nodes. For that purpose, RFnest converts the signals from each transmitter to baseband, processes it with FPGA to impose channel effects, and then converts it back to 2.4~GHz before sending it to the corresponding receiver. 
	
	Our emulation tests for a multi-hop mobile network show that DRL effectively interacts with and learns from the network environment, and builds the best strategy to control the transmit power for WiFi communications. This way, DRL provides major improvements to energy efficiency (up to $24\%$)  and throughput (up to $22\%$) compared to the benchmark schemes of fixed and myopic power allocations.

	The contributions of the paper are summarized below:
	
	\begin{itemize}
		
		\item  A novel DRL-based power control scheme is presented to improve throughput and energy performance of WiFi communications in a multi-hop network of mobile nodes.
		
		\item This algorithmic solution is implemented in an embedded platform that integrates the real network protocol stack with WiFi transceiver for each node.
		
		\item Emulation tests are executed with real RF devices in a controllable and repeatable experimentation environment, such that the proposed solution is compared with the benchmark scheme under the same channel conditions.
		
		\item Performance improvement in terms of throughput and energy efficiency is demonstrated through emulation tests compared to the benchmark schemes of (i) using a fixed transmit power (randomly selected over time) and (ii) solving the optimization myopically.  
		
	\end{itemize}
	
	The rest of the paper is organized as follows. Section~\ref{Sec:Sec2} describes the system model. Section~\ref{Sec:Sec3} presents the DRL-based power control solution. Section~\ref{Sec:Sec4} provides the performance evaluation results. Section~\ref{Sec:Sec5} concludes the paper. 
	
	\section{System Model} \label{Sec:Sec2}
	\subsection{Network Model}
	A multi-hop wireless network of mobile nodes is considered. Each node equipped with a WiFi transceiver (802.11n) communicates with each other (potentially over multiple hops) through WiFi. Nodes are moving according to the random waypoint model. The time is indexed with respect to the WiFi frame duration. At any given time, each node measures channel conditions, namely the link quality and the signal strength, and (if transmitting) adjusts its transmit power over time to adapt to dynamics of network topology and channel as well as dynamics of the network protocol itself. DRL is used to determine the level of transmit power. Two benchmark schemes of fixed and myopic power allocations are considered for comparison purposes.
	
	Each node may act as a source, a destination, or a relay for the multi-hop network traffic. Each node generates random unicast traffic with the randomly selected destination. For multi-hop routing, the Babel routing algorithm \cite{babel} is used as a loop-avoiding distance-vector routing protocol for IPv6 and IPv4 with fast convergence properties. With computationally light implementation, it is suitable for embedded platforms. Expected transmission count (ETX) is used as the link cost. 
	
	\subsection{Embedded Radio Implementation Approach}
	
\begin{figure}[tbh!]
		\centering
		\includegraphics[width=0.9\columnwidth]{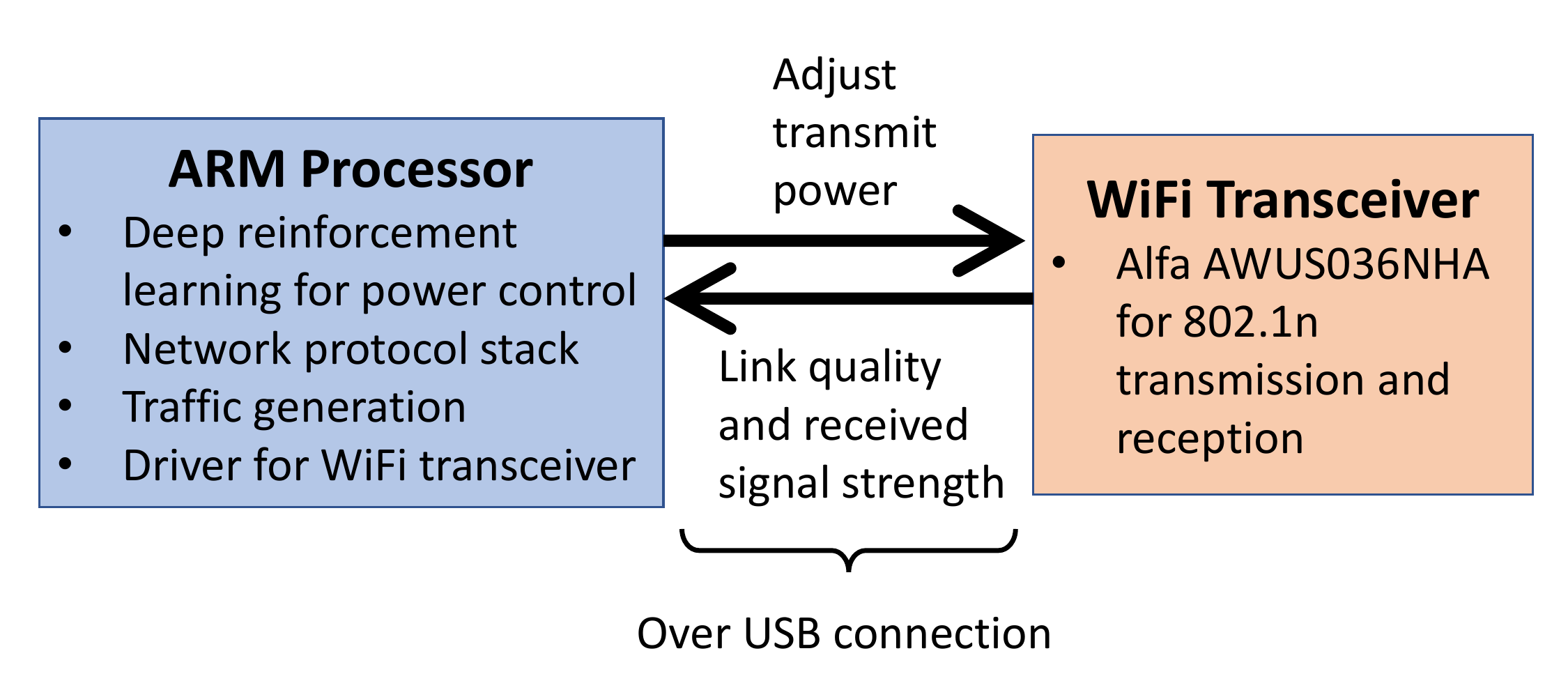}
		\caption{The structure of a node with WiFi transceiver and embedded processor.}\label{fig:Emulation1}
	\end{figure}

	A system-level implementation is considered. For each node, the proposed power control scheme with DRL is implemented on the ARM processor as an embedded solution that controls the Alfa AWUS036NHA USB wireless adapter for 802.11n communications at $2.4$~GHz band. The transmit power ranges from $0$~dBm to $20$~dBm. The routing protocol is also implemented on the ARM for fast operation at the edge device. The structure of a node with WiFi transceiver and embedded processor is shown in Fig~\ref{fig:Emulation1}.
	
	\subsection{Performance Evaluation Approach based on Emulation Tests with Real Transceivers}
	
\begin{figure}[tbh!]
		\centering
		\includegraphics[width=0.9\columnwidth]{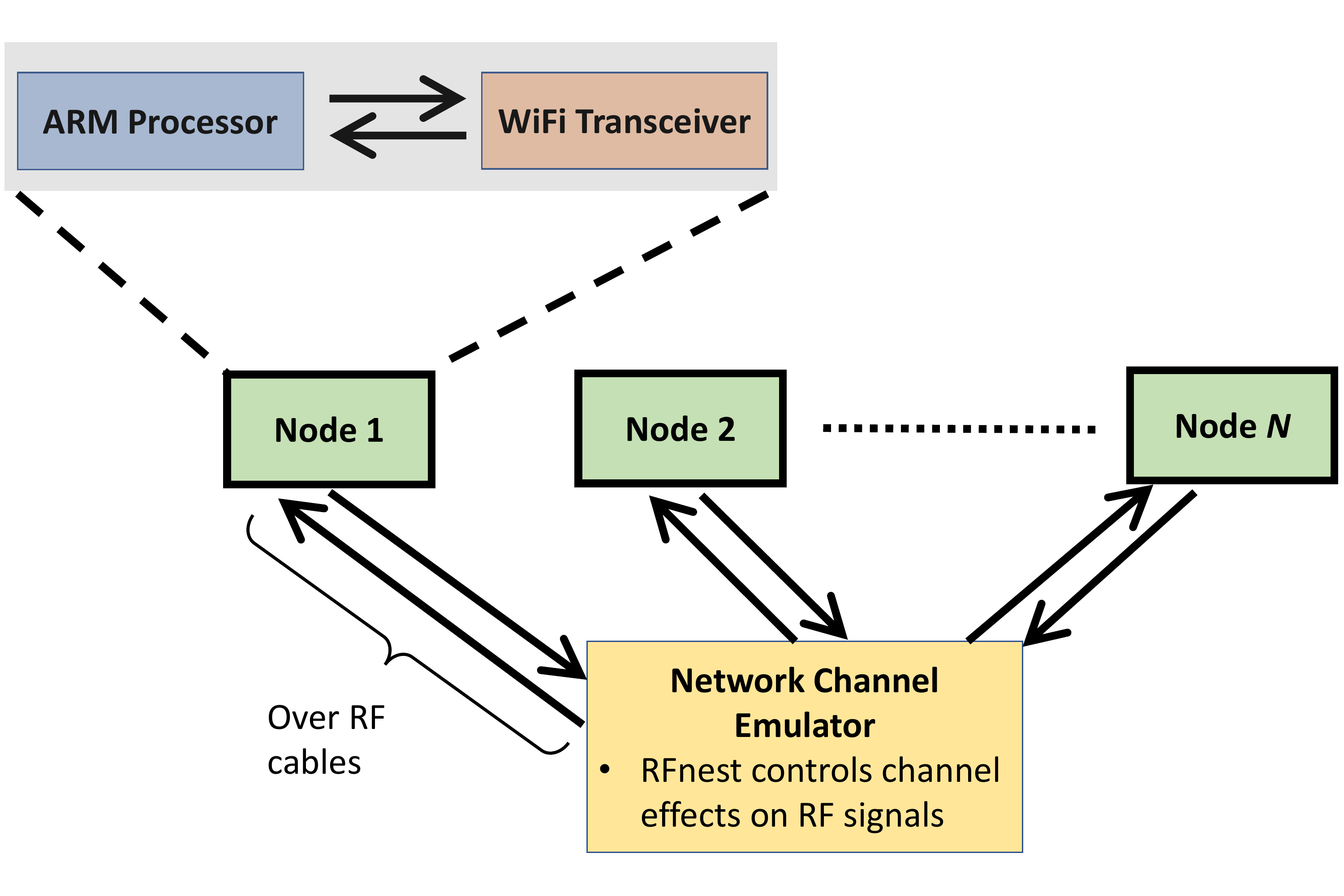}
		\caption{Emulation testbed environment with the network channel emulator controlling channels among nodes corresponding to real transceivers and embedded processors.}\label{fig:Emulation2}
	\end{figure}

	The performance is measured by emulation tests using real WiFi transceivers, real network protocol stack, and real network packet traffic. The emulation environment is illustrated in Fig~\ref{fig:Emulation2}. Instead of over-the-air transmissions, transmissions are made over emulated channels. RFnest is used as the network channel emulator \cite{RFnest}. The antenna of each transceiver is removed and connected via an RF cable to the network channel emulator. Signals out of each transmitting node at 2.4~GHz are fed to the network channel emulator that are converted to the baseband and processed by the FPGA of RFnest to induce the desired channel effects (such as pathloss and multipath), and then converted back to 2.4~GHz and  fed to the antenna ports of the receiver node. 
	
	Channels are controlled by the network channel emulator to reflect the mobility effects. In experiments,  random waypoint model is used for node mobility. A full mesh network channel emulation is considered such that interference effects among nodes are also induced in addition to the channel effects. The emulation of interference effects drives the CSMA/CA-based backoff mechanism of WiFi. These emulation tests allow us to compare the proposed solution with the benchmark scheme under the same channel conditions. 
	
	\section{Deep Reinforcement Learning for Controlling WiFi Power} \label{Sec:Sec3}
	
	Each node adjusts its transmit power at any given time. For that purpose, a DRL solution is used. As the network conditions change over time, DRL provides the mechanism to adapt to the network environment (including network topology, channel, and traffic effects) compared to supervised learning approaches when a model is learned from a predefined labeled dataset. With RL, nodes can interact with the environment, perform actions, and learn over time by trial and error.
	
	In RL, nodes acting as agents take actions and receive rewards from the environment, thereby adapting their decisions in real time to the changes in the environment. However, the computational overhead involving the maintenance of the Q-table grows significantly with the state and action space [7], posing a practical limitation for implementation. DRL reduces the computational burden of conventional RL, leading to a more practical implementation. One way of implementing DRL is based on training of a DQN. The DQN involves a deep neural network to approximate the Q-function in Q-learning. The objective in Q-learning is to maximize the expected discounted reward. The drawback of Q-learning is that the magnitude of Q-value evaluations grows exponentially with the state and action space. Thus, the DQN overcomes this drawback by approximating the Q value with the training of a deep neural network.
	
			%
			%
				%
			%
	%
	%

	Q-learning is formulated as follows. The reward at time $t$ is $r_t$ and the discounted reward is $R_t = r_t + \gamma r_{t+1} + \gamma^2 r_{t+2} + \cdots$, where $\gamma \in [0,1]$ is the discount factor. The Q-function constructs a policy $\pi$ such that $Q_\pi(s,a) = \mathbb{E}[R_t | s_t = s, a_t = a]$, where $a_t$ and $s_t$ denote the action and the state, respectively, at time $t$. The optimal action-value function $Q^\star(s,a) = \max_\pi Q_\pi (s,a)$ obeys the following Bellman equation:
	\begin{align} \label{Bellman}
		Q^\star(s,a) = \mathbb{E}_{s'} \left[ r + \gamma \max_{a'} Q^\star (s',a') | s,a \right],
	\end{align}
	where $s'$ denotes the next state and $a'$ is the action that maximizes the $Q^\star(s',a')$ expression.
	The deep Q-learning uses a deep neural network, parameterized by ${\theta}$ to represent $Q(s,a; {\theta})$, which is iteratively optimized \cite{DQN} by maximizing 
	\begin{align}
		\mathbb{E}_{s,a,r,s'} \left[ r \: | \: \gamma \max_{a'}Q(s',a'; \theta_i^{-}) - Q(s,a; \theta_i) \right], 
	\end{align}
	where $\theta_i$ is the set of parameters for the deep neural network at iteration $i$ and $\theta_i^{-}$ represents the parameters of the target network that is frozen for a number of iterations while updating the value (online) network. In deciding which actions to take, $\epsilon$-greedy policy is used to select the action that maximizes the Q-value with probability $1-\epsilon$ and a random action with probability $\epsilon$ that balances the exploration-exploitation tradeoff. The DQN agent \cite{DQN2} stores the experiences $(s,a,r,s')$ in a replay memory to train the deep neural network by sampling mini-batches of experiences, which is known as experience replay. 
	
	For any node $n$ at time $t$, state $s_n(t)$, action $a_n(t)$, and reward $r_n(t)$ are defined as follows.
	
	\noindent \textbf{\emph{State}} $s_n(t)$: The state corresponds to the tuple of $\{P_n(t), L_n(t), S_n(t) \}$, where $P_n(t)$, $L_n(t)$, and $S_n(t)$ are the transmit power level, the link quality and the signal strength, respectively, for node $n$ at given time $t$. In our experiments, based on the WiFi transceiver properties, $P_n(t)$ takes values from $0$~dBm to $20$~dBm with $1$~dB increments, $L_n(t)$ takes values from $0$ to $70$ with $1$ increments, and $S_n(t)$ takes values from $-110$~dBm and $-40$~dBm with $1$~dB increments.

	\noindent \textbf{\emph{Action}} $a_n(t)$: The action of node $n$ at any given time $t$ is changing the transmit power, namely selecting $\Delta P_n(t) = P_{n}(t)-P_{n}(t-1)$, where $\Delta P_n(t) \in \{\Delta, 0, - \Delta\}$. There are three possible actions. In experiments, we set $\Delta = 1$ (measured in dBm).
	
	\noindent \textbf{\emph{Reward}} $r_n(t)$: The reward of node $n$ at time $t$ is computed as 
	\begin{equation}
		r_n(t) =  \frac{T_n(t)}{E_n(t)} - c \cdot \Delta P_n(t),
	\end{equation}
	where $T_n(t)$ is the throughput received by node $n$ at time $t$, $E_n(t)$ is the total energy consumption normalized with respect to the packet duration that corresponds to one time increment, and $c$ is a positive constant ($c = 0.1$ in our experiments). The unit for throughput is Mbps and the unit of energy efficiency is Mbps/Joule. The second term in the reward corresponds to the penalty of changing the transmit power (it is needed to avoid excessive fluctuations in decision space). When node $n$ is not transmitting at time $t$, $r_n(t) = 0$ is imposed.  
	
	Note that if conventional Q-learning is used, the corresponding size of state space is $21 \times 71 \times  71 = 105,861$ (there are $21$ levels for the transmit power, $71$ levels for the link quality, and $71$ levels for the signal strength). As the number of possible actions is $3$, the size of Q-table in case of conventional RL would be $105,861 \times 3 = 317,583$. It is practically difficult (if not infeasible) to maintain and operate over such a large Q-table at an edge device that relies on limited memory and processing capabilities. Hence, a DRL approach is adopted to reduce the computational complexity and memory requirements for the support of edge applications. 
	
	To approximate the Q value in DQN, a deep neural network is trained. The deep neural network maps the input states to the pairs of actions and Q-values. For that purpose, a three-layer feedforward neural network (FNN) is used. The input layer takes the states as the input. There are 140 and 70 neurons in the hidden layers. The output layer has 3 neurons corresponding to three possible actions. A rectifying linear unit (ReLU) activation function is used for the hidden layers. The FNN is trained by minimizing the mean squared error (MSE) loss function given by
	\begin{align}\label{eq:mse}
		\ell(\theta_i) = \left(r_n(t) +
		\gamma \max_{a'} Q^\star (s',a',\theta_i^-) - Q(s,a,\theta_i)\right)^2 .
	\end{align}
	
\begin{figure}[tbh!]
		\centering
		\includegraphics[width=0.9\columnwidth]{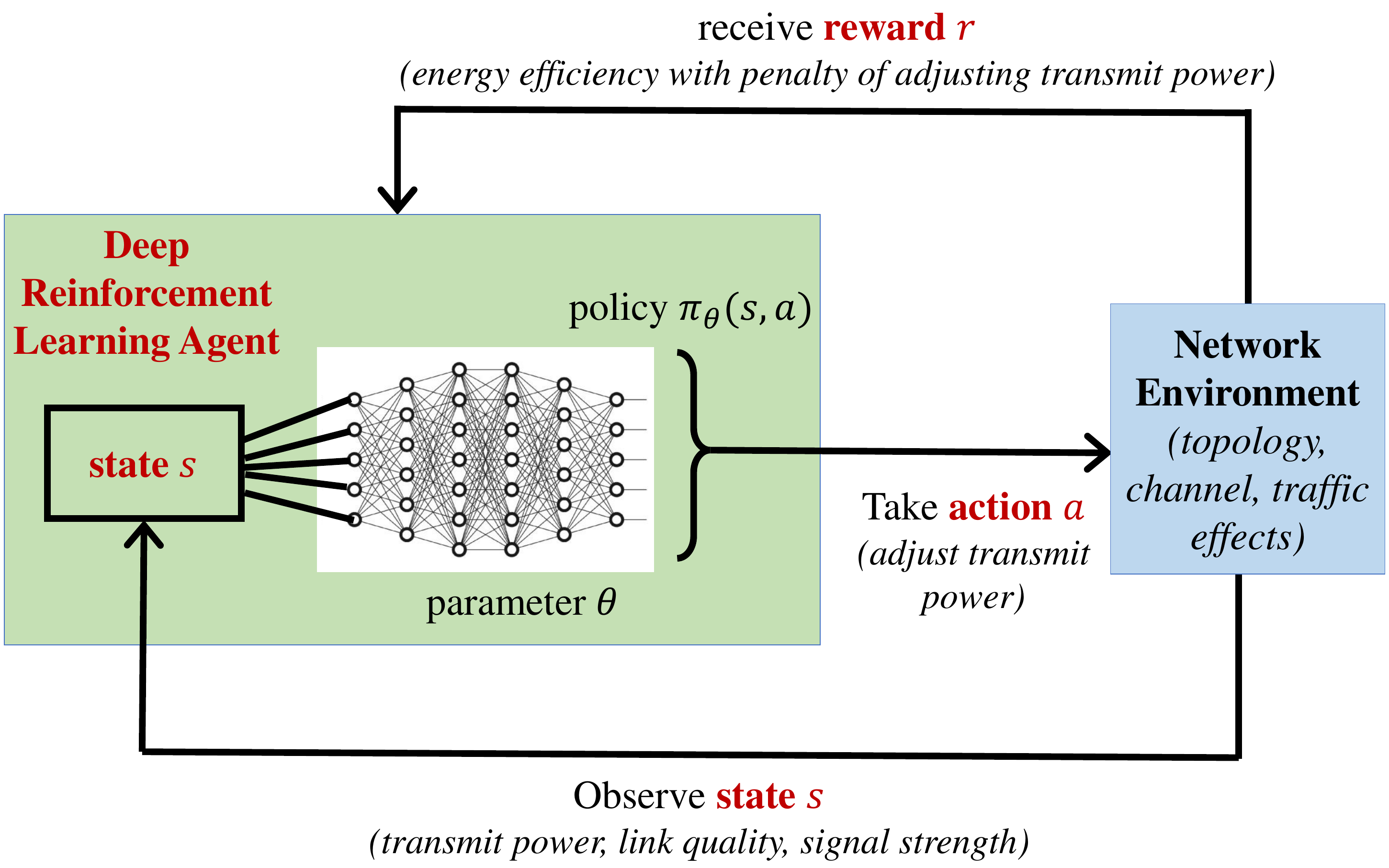}
		\caption{Illustration of the overall DRL solution.}\label{fig:RL}
	\end{figure}

	The overall DRL solution is summarized in Fig.~\ref{fig:RL}. The DRL agent for each node is trained by the help of its neighbors in a distributed setting without relying on a centralized controller. Each node observes its state and acquires states of other nodes from local messages to form its observation vector $\boldsymbol{s}_i(t)$, selects and performs its action, receives reward $r_i(\boldsymbol{s}_i(t), a_i(t))$, broadcasts its action $a_i(t)$ to its neighbors, stores transition in replay memory, and updates weights for its DQN. The details of the deep Q-learning algorithm are given in Algorithm~\ref{DQN}.
	
	\begin{algorithm}[!]
		\small
		\SetAlgoLined
		Initialize replay memory $\mathcal{M}$ \;
		Initialize action-value function $Q$  with random weights\;
		\For{episode = $1,...,M$ }{
			
			\For{episode = $1,...,M$ }{
				\For{each node $i$ (in parallel)}
				{
					\begin{enumerate}
						\item Observe its state and acquire states of other \\  nodes from local messages to form observation \\
						vector  $\boldsymbol{s}_i(t)$.
						\item Select action 
						\\$a_i(t)=\arg\max_{a_i(t) \in \mathcal{A}} Q(\boldsymbol{s}_i(t), a_i(t))$ \\ w.p. $1-\epsilon$ or a random action $a_i(t) \in \mathcal{A}$ w.p. $\epsilon$.
						\item Perform an action $a_i(t)$.
						\item Receive reward $r_i(\boldsymbol{s}_i(t), a_i(t))$. 
						\item Broadcast its action $a_i(t)$ to its neighbors.
						\item Store transition in replay memory $\mathcal{M}$.
						\item Update weights. 
					\end{enumerate}
					
				}
				
			}
		}
		\caption{Deep Q-learning algorithm.}\label{DQN}
	\end{algorithm}
	\normalsize


		\begin{figure}[tbh!]
			\centering
			\includegraphics[width=0.8\columnwidth]{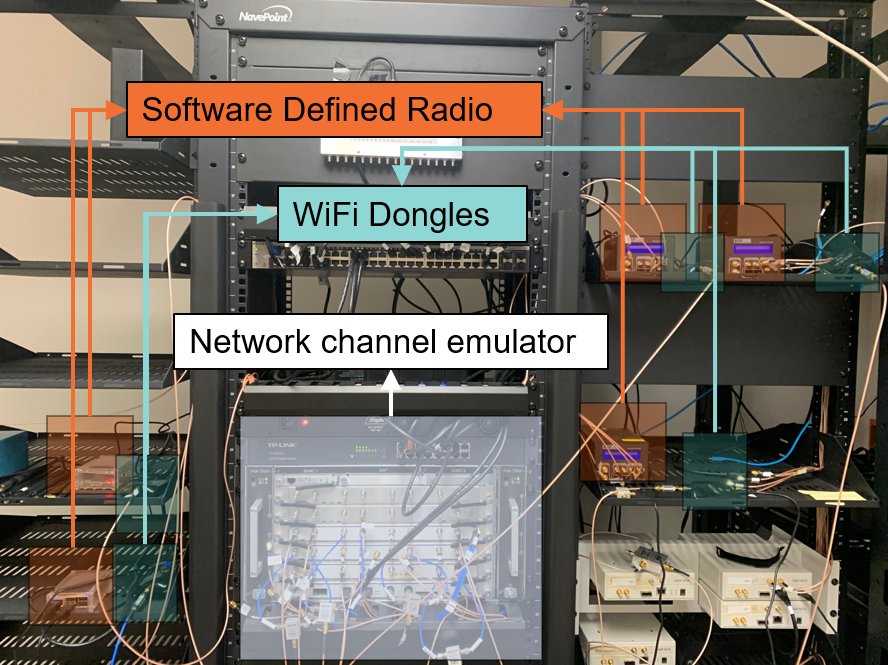}
			\caption{The testbed for hardware-in-the-loop emulation tests.}\label{fig:lab_setup}
		\end{figure}
		
		\section{Performance Evaluation with Emulation Tests} \label{Sec:Sec4}
		For system-level evaluation, emulation tests are performed with embedded platforms, each (corresponding to one node) equipped with an Alfa AWUS036NHA USB wireless adapter for 802.11n and an ARM processor hosted on a software-defined radio (SDR). The ARM processor hosts and executes the full network protocol stack including the WiFi transmit power control based on DRL.

		The emulation test setup is shown in Fig.~\ref{fig:lab_setup}.
		The frequency of communications is set as 2.4~GHz. The power control allows each node to adjust its transmit power automatically from $0$~dBm to $20$~dBm. Algorithm 1 involves distributed operation of nodes, each computing their actions and exchanging information with neighbor nodes. Therefore, it is independent of the number of nodes and network mobility. A multi-hop mobile network of $5$ nodes is considered in emulations tests, where each node moves according to the random waypoint model with $5$ m/s speed. Network channel emulator (RFnest) imposes the changes in the network topology on channel conditions and controls the RF signals accordingly by accounting for the distances among nodes that change over time. The performance is measured in terms of two metrics:
		\begin{itemize}
			\item \emph{Throughput}: total number of bits received by all destinations per second).
			\item \emph{Energy efficiency}: throughput per unit energy consumption, where energy is measured by total (transmit and processing) power consumption over the packet duration.
		\end{itemize}

\begin{figure} [tbh!]
		\centering
		\includegraphics[width=0.88\columnwidth]{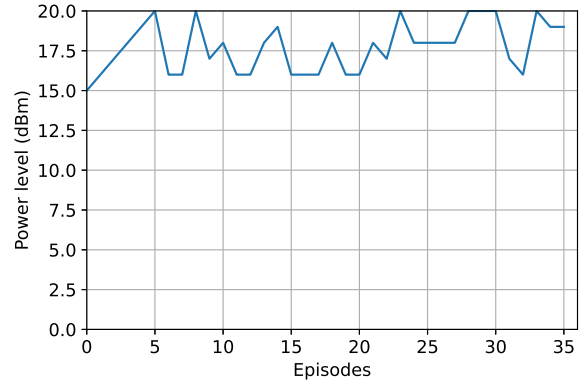}
		\caption{Transmit power of DRL.}\label{fig:drl_power}
	\end{figure}

\begin{figure} [tbh!]
		\centering
		\includegraphics[width=0.86\columnwidth]{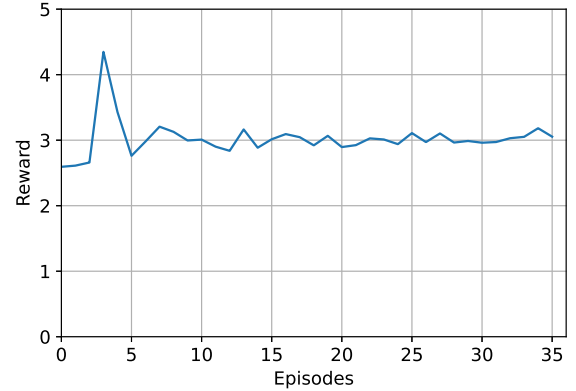}
		\caption{Reward of DRL.}\label{fig:drl_reward}
	\end{figure}

	Network traffic is generated with iPerf3. The throughput by each node is measured and energy efficiency is computed. In this setting, the node that is training becomes the iPerf client and neighbors become the iPerf servers. This approach makes training more communication-efficient. Fig.~\ref{fig:drl_power} shows an example of how 
	transmit power levels change over time as determined by the DRL agent of a node that reacts to changes in the network (including topology, channel and traffic effects) and adjusts the transmit power levels accordingly. Each episode corresponds to one MAC frame. In 802.11n, the frame size is bounded by a limit of about $5.5$ ms.  The reward of the corresponding DRL agent is shown in Fig.~\ref{fig:drl_reward} as a function of time. Note that the DRL environment considered in this paper differs from most RL environments, where the maximum reward is a constant. Since the link distances change, so does the maximum reward that an agent can achieve in the formulation considered in this paper.

	%
	%
	
		
		
	
	\begin{table} [tbh!]
		\caption{Average performance measured in emulation tests.}
		\centering
		
		\begin{tabular}{c|c|c}
			\toprule
			Algorithm  & Energy Efficiency  & Throughput \\ 
			& (Mbps/J)				& (Mbps) \\
			\midrule
			Fixed power & $2.601$ & $5.723$  \\   
			Myopic allocation & $2.893$  & $5.983$ \\ 
			\textbf{DQN}  & $\textbf{3.232}$ & $\textbf{6.964}$ \\
			\bottomrule				
		\end{tabular}
		
		\label{table:resource}
	\end{table}
	
	The average performance of the proposed scheme (based on DRL) collected over more than 30 epochs is shown in Table~\ref{table:resource}. Two benchmark schemes are considered: (i) random selection of transmit power from $0$~dBm, $10$~dBm, or $20$~dBm and (ii) myopic allocation of transmit power (by solving (\ref{Bellman}) with $\gamma = 0$). Experiments with real radios and embedded implementation show that DRL achieves up to $24\%$ performance gain in terms of energy efficiency and $22\%$ performance gain in terms of throughput.


	The DRL model that is used by the ARM processor takes only 20.8~KFlops to make a decision, which is fairly small compared to the models typically used in computer vision. The ARM processor (such as the one available on the Raspberry Pi-4B (1GB)) consumes 1.50 GFlops/W \cite{gflop}. Then, the power control algorithm takes only 13.9~$\mu W$ (or -18.6 dBm), which is fairly small compared to the amount of power required to transmit or receive. Also, power control updates can be made less frequently than transmit decisions, further reducing the power consumption for computing.


\section{Conclusion} \label{Sec:Sec5}
A DRL solution for transmit power control was presented for WiFi network systems. In a multi-hop mobile network, each node observes the states of transmit power, link quality and signal strength, takes actions of selecting its transmit power, and optimizes a cumulative reward based on energy efficiency following a distributed operation. For each node, this solution was implemented on an embedded platform that integrates an ARM processor for the network protocol stack (including power control with DRL) and a WiFi transceiver. High-fidelity emulation tests with real packet traffic showed that the DRL-based power control can achieve up to $24\%$ improvement in energy efficiency while maintaining $22\%$ improvement in throughput compared to the fixed and myopic power allocation schemes, thereby prolonging network lifetime as well as serving the ever-increasing demand for high communication rates.

\end{document}